\documentclass[12pt]{iopart} 

\begin{document} 

\jl{3}

\title[Direct and indirect exciton mixing in a slightly asymmetric DQW]
{Direct and indirect exciton mixing in a slightly asymmetric double quantum well}

\author{Francisco Vera}

\address{Universidad Cat\'{o}lica de Valpara\'{\i}so, Av. Brasil 2950, Valpara\'{\i}so,
Chile}

\begin{abstract}
We studied, theoretically, the optical absorption spectra for a slightly asymmetric
double quantum well (DQW), in the presence of electric and magnetic fields.
Recent experimental results for a 10.18/3.82/9.61 nm GaAs Al\( _{.33} \)Ga\( _{.67} \)As
DQW show clearly the different behavior in the luminescence peaks for the indirect
exciton \( IX \) and left direct exciton \( DX \) as a function of the external
electric field. We show that the presence of a peak near the \( DX \) peak,
attributed to an impurity bound left \( DX \) in the experimental results,
could be a consequence of the non-trivial mixing between excitonic states.
\end{abstract}






\submitted

\section{Introduction}
\label{sec:Introduction}

In this work, we studied the excitonic energy levels in an asymmetric vertical
double quantum well 
\cite{kamizato}-\cite{bastard} as a function of the magnetic and electric
field strengths. Within the effective-mass approach we expanded the excitonic
wave function in an orthogonal basis formed by products of electron and hole
wave functions along the crystal growth direction \( z \), and one-particle
solutions of the magnetic Hamiltonian in the \( x \)-\( y \) plane. The Coulomb
potential between electrons and holes produces off-diagonal terms thus mixing
our basis states. We obtained the energy spectra and wave functions by diagonalizing
the excitonic Hamiltonian in a truncated basis. We applied our method to study
the excitonic states in a GaAs Al\( _{.33} \)Ga\( _{.67} \)As DQW, for the
specific case of a DQW composed of a left well of \( 10.18 \) nm, a barrier
of \( 3.82 \) nm, and a right well of \( 9.61 \) nm. The effects of external
electric and magnetic fields on the luminescence (PL) intensity for this heterostructure
has been recently studied experimentally by Krivolapchuk et al.\cite{krivolapchuk1999},
\cite{krivolapchuk1998}.

\maketitle

\section{Formalism}

\label{sec:Formalism}

The effective-mass Hamiltonian for excitons in a double quantum well in the
diagonal approximation \cite{bastard} and in the presence of a magnetic field
pointing towards \( z \), can be written as 
\begin{equation}
\label{h-total}
H=H_{0}(z_{e})+H_{0}(z_{h})+H_{mag}(r)+V_{coul}(r,|z_{e}-z_{h}|),
\end{equation}
 where \( H_{0}(z_{e}) \) is the one-dimensional Hamiltonian for electrons,
\begin{equation}
\label{h-ele}
H_{0}(z_{e})=p_{ze}^{2}/2m_{ze}+V_{e}(z_{e}),
\end{equation}
 \( H_{0}(z_{h}) \) is the one-dimensional Hamiltonian for holes, 
\begin{equation}
\label{h-hole}
H_{0}(z_{h})=p_{zh}^{2}/2m_{zh}+V_{h}(z_{h}),
\end{equation}
 and \( V_{e}(z_{e}) \) (\( V_{h}(z_{h}) \)) is the potential that defines
the double quantum well for electrons (holes) in the five regions of \( z \).
We included the electric field in \( V_{e}(z_{e}) \) and \( V_{h}(z_{h}) \)
by a shift in the potential in stair steps similar to Fig.\ \ref{pot-z}. \( H_{mag}(r) \)
is the magnetic Hamiltonian in the symmetric gauge, which depends on the relative
coordinates of electrons and holes in the \( x-y \) plane, 
\begin{equation}
\label{h-mag}
H_{mag}=\frac{(\vec{p}-q\vec{A})^{2}}{2\mu }+\frac{qB}{m_{h,x-y}}l_{z},
\end{equation}
 where \( \vec{p} \), \( \mu  \) and \( m_{h,x-y} \) are the momentum operator,
reduced mass, and hole mass respectively, defined on the \( x-y \) plane.

\( V_{coul}(r,|z_{e}-z_{h}|) \) is the Coulomb potential between electrons
and holes, including an effective dielectric constant for the system.

We expanded the solutions of the Hamiltonian (\ref{h-total}), as a linear combination
of products of eigenfunctions of the magnetic Hamiltonian in the \( x-y \)
plane (\ref{h-mag}), and eigenfunctions of the electron and hole Hamiltonians
in the \( z \) direction {[}(\ref{h-ele}) and (\ref{h-hole}){]}, 
\begin{equation}
\label{expansion}
\Psi ^{exc}_{n}=\sum _{\nu _{r},\nu _{e},\nu _{h}}C^{n}_{\nu _{r},\nu _{e},\nu _{h}}\psi _{\nu _{r}}(r,\phi )\psi _{\nu _{e}}(z_{e})\psi _{\nu _{h}}(z_{h}),
\end{equation}
 in which, in the symmetric gauge,
\begin{eqnarray}
\psi _{\nu _{r},l}= & \frac{1}{2\pi }\left( \frac{2(n-l/2-|l|/2)g_{B}^{|l|+1}}{(|l|/2+n-l/2)!}\right) ^{1/2}e^{il\phi }(\frac{r}{i})^{|l|} & \label{f-mag} \\
 & \times e^{-g_{B}r^{2}/2}L^{|l|}_{n-l/2-|l|/2}(g_{B}r^{2}),\nonumber 
\end{eqnarray}
 where \( g_{B}=\frac{qB}{2\hbar } \), and only \( l=0 \) functions are considered.
The electron wave functions defined in the five regions of \( z \), shown in
Fig.\ \ref{pot-z}, are given in terms of \( sin(z_{e}) \) , \( cos(z_{e}) \),
and \( \pm exp(z_{e}) \). The hole wave functions are given by similar expressions.

The Coulomb interaction produces off-diagonal terms by mixing our basis states.
In order to obtain the system of equations for the coefficients in expansion
(\ref{expansion}), we need to evaluate the Coulomb integrals
\begin{equation}
\label{i-total}
\int d\phi drdz_{e}dz_{h}\psi ^{*}_{\nu' _{r}}\psi _{\nu' _{e}}\psi _{\nu' _{h}}V_{coul}(r,|z_{e}-z_{h}|)\psi _{\nu _{r}}\psi _{\nu _{e}}\psi _{\nu _{h}}.
\end{equation}

The \( \phi  \) integral is trivial, because of \( l_{z} \) conservation.
Using the explicit expansion of Laguerre Polynomials (\( L_{n} \)) in \( \psi _{\nu' _{r}} \)
and \( \psi _{\nu _{r}} \) (\ref{f-mag}) and after solving the \( r \) integral,
eqn.~(\ref{i-total}) can be written as a sum of terms of the form 
\begin{equation}
\label{i-final}
\int _{0}^{\infty }d\alpha (\frac{\alpha ^{2}}{4g_{B}})^{m}e^{\frac{-\alpha ^{2}}{4g_{B}}}\int _{-\infty }^{\infty }dz_{e}dz_{h}\psi _{\nu' _{e}}\psi _{\nu' _{h}}e^{-|z_{e}-z_{h}|\alpha }\psi _{\nu _{e}}\psi _{\nu _{h}}
\end{equation}

The \( z_{e} \)- and \( z_{h} \)- integrals can be solved analytically. The
evaluation of these integrals is cumbersome due to the large number of terms,
resulting from the five different regions of the potential. The \( z_{e} \)-
and \( z_{h} \)- integrals contain both decoupled terms in which the \( z_{e} \)-
and \( z_{h} \)- integrals are independent of each other and coupled terms
where the integration limits of the \( z_{h} \)-integral contains \( z_{e} \).
The remaining \( \alpha  \) integral must be calculated numerically.

By diagonalizing the system of equations resulting for the coefficients in expansion
(\ref{expansion}), in a truncated basis, we obtained the energies and wave
functions for the first excitonic levels. Evaluating the oscillator strength
we obtained the optical absorption spectra.

\section{Results}

We calculated the excitonic energy levels for a slightly asymmetric 10.18/3.82/9.61
nm GaAs Al\( _{.33} \)Ga\( _{.67} \)As double quantum well, in the presence
of electric and magnetic fields. The band gap used in our calculations is given
by \( E_{g}(x)=1.52+1.36x+0.22x^{2} \) (\( x=0.33 \)). The band-gap offset
considered was \( 60\% \) for the conduction band and \( 40\% \) for the valence
band. For all five regions in the double quantum well, we used the same electronic
mass \( m_{e}=0.067m_{0} \) , the \( x-y \) plane heavy-hole mass, \( m_{hh,x-y}=0.1m_{0} \),
the \( z-axis \) heavy-hole mass \( m_{hh,z}=0.45m_{0} \) , the light-hole
masses \( m_{lh,x-y}=0.2m_{0} \) and \( m_{lh,z}=0.08m_{0} \), and a dielectric
constant \( \epsilon =12.5\epsilon _{0} \). In our calculations we used a truncated
basis set composed of \( 12 \) Landau wave functions, four electronic wave
functions, five heavy-hole wave functions, for which most of our results converged
to less than a tenth of meV. For magnetic fields below \( 5 \) T, we would
need to increase the number of Landau levels considered. We included a fictitious
width of \( 1.5 \) meV for each optical absorption peak.

\subsection{Energies and wave functions for electrons and holes}

\label{sec:Energies}

In this section we present our results for the energies and wave functions of
electron and holes in a one-dimensional DQW, as a function of the external electric
field. We are interested in regions of electric field where two different electron
(or hole) energy levels cross each other leaving a gap of a few meV between
them. When the wave functions involved in this kind of anticrossing are important
for the optical absorption, we expect to find interesting behaviors in the peaks
of the excitonic optical absorption.

Figure \ref{E_F} shows the electron (upper plot) and hole (lower plot) energy
levels for this DQW as a function of the external electric field. In the range
from \( 0 \) to \( 60 \) kV/cm (not shown in the plot), there are crossings
in the hole energy levels near \( 5 \), \( 15 \), \( 25 \) and \( 40 \)
kV/cm. These crossings do not met the two conditions explained before, and will
not produce a noticeable behavior in the excitonic optical absorption. In the
range from \( 60 \) to \( 85 \) kV/cm there is a region of electric fields
near \( 67 \) kV/cm, indicated with an arrow, where the second and the third
electronic energy levels anticross. To understand the excitonic optical absorption
it is necessary to know the behavior of the electron and hole DQW wave functions
for electric field below, inside, and above the anticrossing region.

Figure \ref{WF_0} shows the first four electron (upper plot) and hole (lower
plot) wave functions for zero external electric field. Numbers indicate the
order of the energy levels, and the most important states for the excitonic
optical absorption are represented by a solid line. The first state mostly localized
in the left (wider) well is called \textbf{L1} and the first state mostly localized
in the right well is called \textbf{R1}. It is clear from this figure that holes
are almost completely localized in the left or right wells, and that electrons
have strong components over both wells. This different behavior for electrons
and holes is a consequence of the effective mass differences for electron and
holes and the size of the wells and barrier in this sample.

Figure \ref{WF_71} shows the electron (upper plot) and hole (lower plot) wave
functions for an external electric field of \( 71 \) kV/cm, where the most
important states for the excitonic optical absorption are represented by a solid
line. Comparing with the previous case of electrons in zero electric field,
we note that: the first level \textbf{L1} get much more localized in the left
well, the second level \textbf{R1} changes dramatically into \textbf{L2}, and
the third level behaves as \textbf{R1}. For holes we can see that the electric
field forces the wave functions to be localized in the right well, and the fourth
level is localized in the left well (corresponding to \textbf{L1}).

Figure \ref{WF_85} shows the electron (upper plot) and hole (lower plot) wave
functions for an external electric field of \( 85 \) kV/cm. Again the most
important states for the excitonic optical absorption are represented by solid
lines. Comparing with the previous case of a \( 71 \) kV/cm electric field,
we note that electron wave functions remains almost unchanged, and that the
fifth hole wave function behaves as \textbf{L1}.

\subsection{Optical absorption as a function of the electric field}

\label{sec:Electric}

Increasing the electric field in a double quantum well produces a shift in the
electron and hole energy levels and a change in the localization of the respective
wave-functions. The combination of both effects changes the behavior of the
excitonic states and affects, in a non trivial way, the optical absorption spectrum
of these systems.

Figure \ref{Abs_F} shows the electric-field effects on the optical absorption
spectra for an external magnetic field of \( 10 \) T. Each curve, corresponding
to a different value of the electric field, has been displaced for clarity.
For an electric field of \( 60 \) kV/cm (upper curve) there are two peaks in
the excitonic optical absorption. The first peak corresponds to a direct exciton
\textbf{L1L1}, where the constituent electron and hole are localized in the
left (wider) well. The second peak corresponds to a direct exciton \textbf{R1R1},
where the electron and hole are localized in the right well. For an electric
field of \( 85 \) kV/cm (lower curve) the same two peaks occur, corresponding
again to the left direct exciton \textbf{L1L1} and the right direct exciton
\textbf{R1R1}. The main difference between the \( 60 \) kV/cm and the \( 85 \)
kV/cm case is the exciton composition in terms of the base states. This different
composition of base states can not be observed experimentally in the previous
limiting cases, but has a strong influence in the optical absorption behavior
for electric fields between \( 60 \) kV/cm and \( 85 \) kV/cm. The complex
structure in the optical absorption spectra in this range of electric fields
can be explained by the formation of excitonic states, which have a strong component
of indirect exciton states. In this case the indirect exciton state corresponds
to \textbf{L2R1}, which is formed by an electron in the second left state and
a hole in first right state. The composition in terms of our base states can
be obtained from figures \ref{WF_0}, \ref{WF_71}, and \ref{WF_85}. When comparing
this figure with Fig. 2 of \cite{krivolapchuk1999}, we see a similar behavior
in the PL peaks. This suggests that the peak attributed to an impurity-bound
direct exciton in the experimental results, could be interpreted as a mixed
state with direct and indirect exciton components. Quantitative agreement between
our results and those in the experimental work is not possible because we included
a strong magnetic field. Peaks originating from light holes states appear at
the right end of this figure (not shown).

\subsection{Optical absorption as a function of magnetic field}

\label{sec:Magnetic}

Increasing the magnetic field in a double quantum well produces a shift in the
excitonic energy levels towards higher energies and an increase in the interaction
energy. The increase in the interaction energy between electrons and holes is
a consequence of two effects: First, the wave function confinement in the \( x \)-\( y \)
plane produces a stronger interaction along the \( z \) direction, as the wave
functions penetrate more into the barrier. Second, the interaction between electrons
and holes in the same plane increases, because their wave functions are confined
to a smaller region.

Figure \ref{Abs_B:F=3D71} shows the magnetic-field effects on the optical absorption
spectra for an external electric field of \( 71 \) kV/cm. The differences in
the interaction energies in the layer plane and along the \( z \) \( axis \)
, when the magnetic field is varied, produces a different behavior in the optical
absorption peaks with direct or indirect exciton components. This difference
produces the splitting of the first peak for a magnetic field of \( 5 \) T
into two peaks for stronger magnetic fields.

Figure \ref{Abs_B:F=3D78} shows the magnetic-field effects on the optical absorption
spectra for an external electric field of \( 78 \) kV/cm. In this case the
magnetic field produces no mixing between excitonic states, and produces different
shifts in energies for direct excitons (\textbf{L1L1} and \textbf{R1R1}) and
the indirect exciton (\textbf{L2R1}). This effect is clearly observed as a function
of magnetic field.

\section{Conclusion}

\label{sec:Conclusion}

In this work we studied the energies and wave functions of excitonic states
in a 10.18/3.82/9.61 nm GaAs Al\( _{.33} \)Ga\( _{.67} \)As double quantum
well, in a magnetic field pointing along the growth direction \( z \). We calculated
the optical absorption spectra as a function of the external electric and magnetic
fields and studied the behavior in the PL peaks for the indirect exciton \( IX \)
and direct exciton \( DX \). We compared our results and those in the experimental
work of \cite{krivolapchuk1999} and found similar behavior in the PL peaks.
This suggests that the peak attributed to an impurity-bound direct exciton in
the experimental results, could be interpreted as a mixed state with direct
and indirect exciton components. 

In our method of calculation we used an orthogonal basis that involves functions
of single-particle solutions of the double quantum well in \( z \) direction,
which makes our method appropriate for small and medium barrier widths. The
study of these systems for magnetic fields below \( 5 \) Tesla within this
method requires to increase the number of Landau levels in our base states,
making the numeric computation excessively time consuming.

\ack

This work was supported by the Fondecyt postdoctoral project No. 3990051 and
by the ``Catedra Presidencial en Ciencias 1998, Francisco Claro''.

\newpage


\Figures

\begin{figure} 

\caption{Potential profile along the \(z\) direction for electrons and holes.
} 

\label{pot-z}

\end{figure}

\begin{figure} 

\caption{One dimensional electron and hole energy levels in a 10.18\-/3.82\-/9.61 nm GaAs/Al\( _{.33} \)Ga\( _{.67} \)As/GaAs
DQW as a function of the electric field strength.} 

\label{E_F}

\end{figure}

\begin{figure} 

\caption{One dimensional electron and hole wave functions in the DQW for zero electric field. The upper plot corresponds to electrons and the lower plot to holes, numbers represent the order in energy for the levels, and letter L (R) indicates that the state is mostly localized in the left (right) quantum well.} 

\label{WF_0}

\end{figure}

\begin{figure} 

\caption{One dimensional electron and hole wave functions in the DQW for an electric field of 71 kV/cm.The upper plot corresponds to electrons and the lower plot to holes, numbers represent the order in energy for the levels, and letter L (R) indicates that the state is mostly localized in the left (right) quantum well.} 

\label{WF_71}

\end{figure}

\begin{figure} 

\caption{One dimensional electron and hole wave functions in the DQW for an electric field of 85 kV/cm.The upper plot corresponds to electrons and the lower plot to holes, numbers represent the order in energy for the levels, and letter L (R) indicates that the state is mostly localized in the left (right) quantum well.} 

\label{WF_85}

\end{figure}

\begin{figure} 

\caption{Optical absorption spectra for this DQW for a 10 Tesla magnetic field and electric fields ranging from 60 kV/cm (upper curve) to 85 kV/cm (lower curve). Labels indicate the main composition of the electron-hole excitonic pair, letter L (R) indicates that the electron or hole state is mostly localized in the left (right) quantum well.} 

\label{Abs_F}

\end{figure}

\begin{figure} 

\caption{Optical absorption spectra for this DQW for a 71 kV/cm electric field and the magnetic field ranging from 5 T (upper curve) to 14 T (lower curve). Labels inside parenthesis indicate a small contibution from these states to the total excitonic wave function.} 

\label{Abs_B:F=3D71}

\end{figure}

\begin{figure} 

\caption{Optical absorption spectra for this DQW for a 78 kV/cm electric field and the magnetic field ranging from 5 T (upper curve) to 14 T (lower curve). } 

\label{Abs_B:F=3D78}

\end{figure}

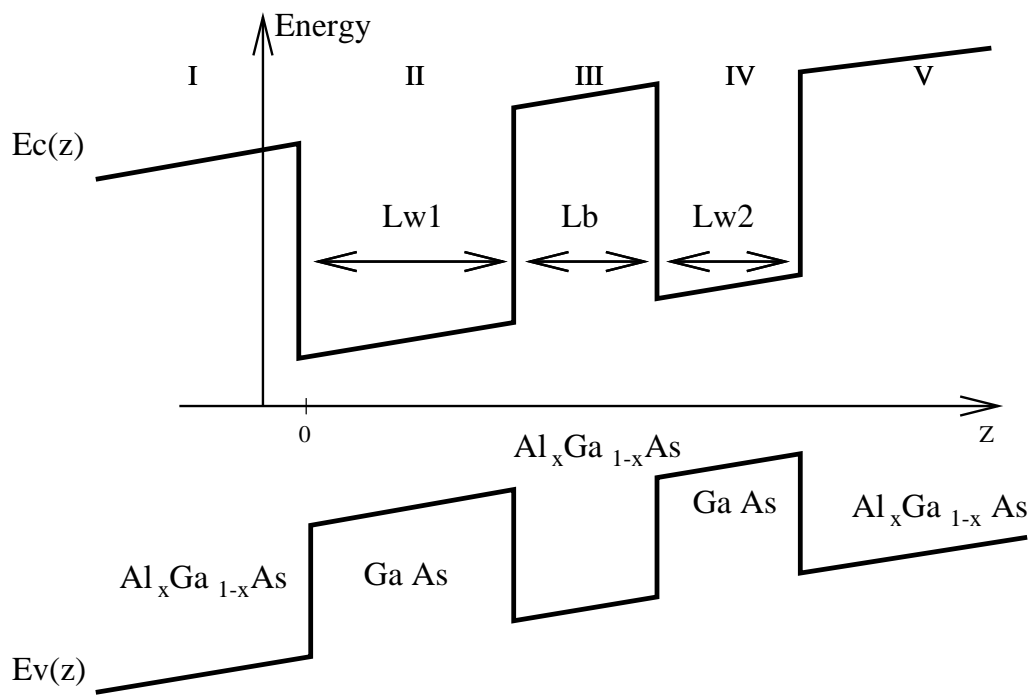
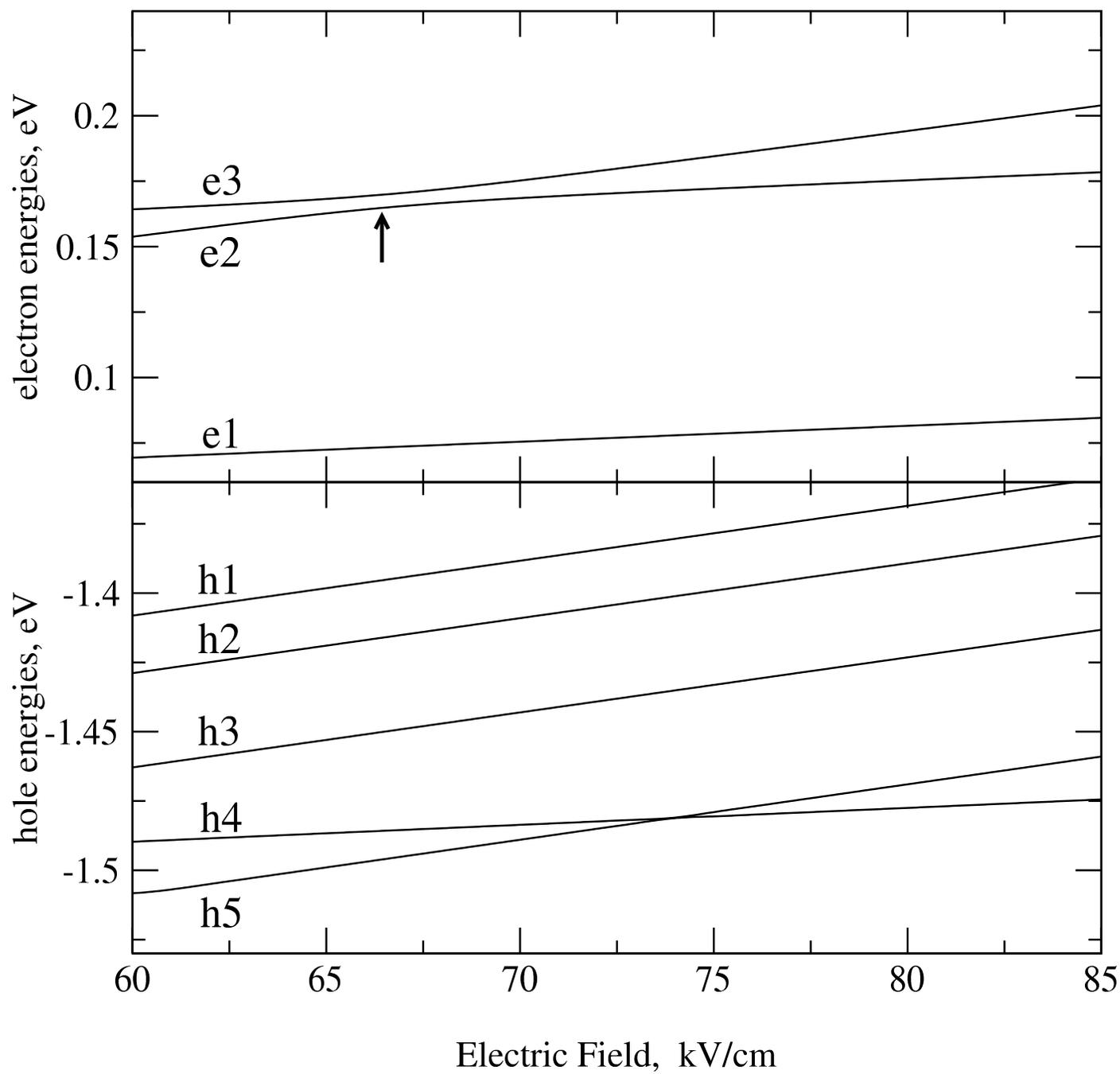
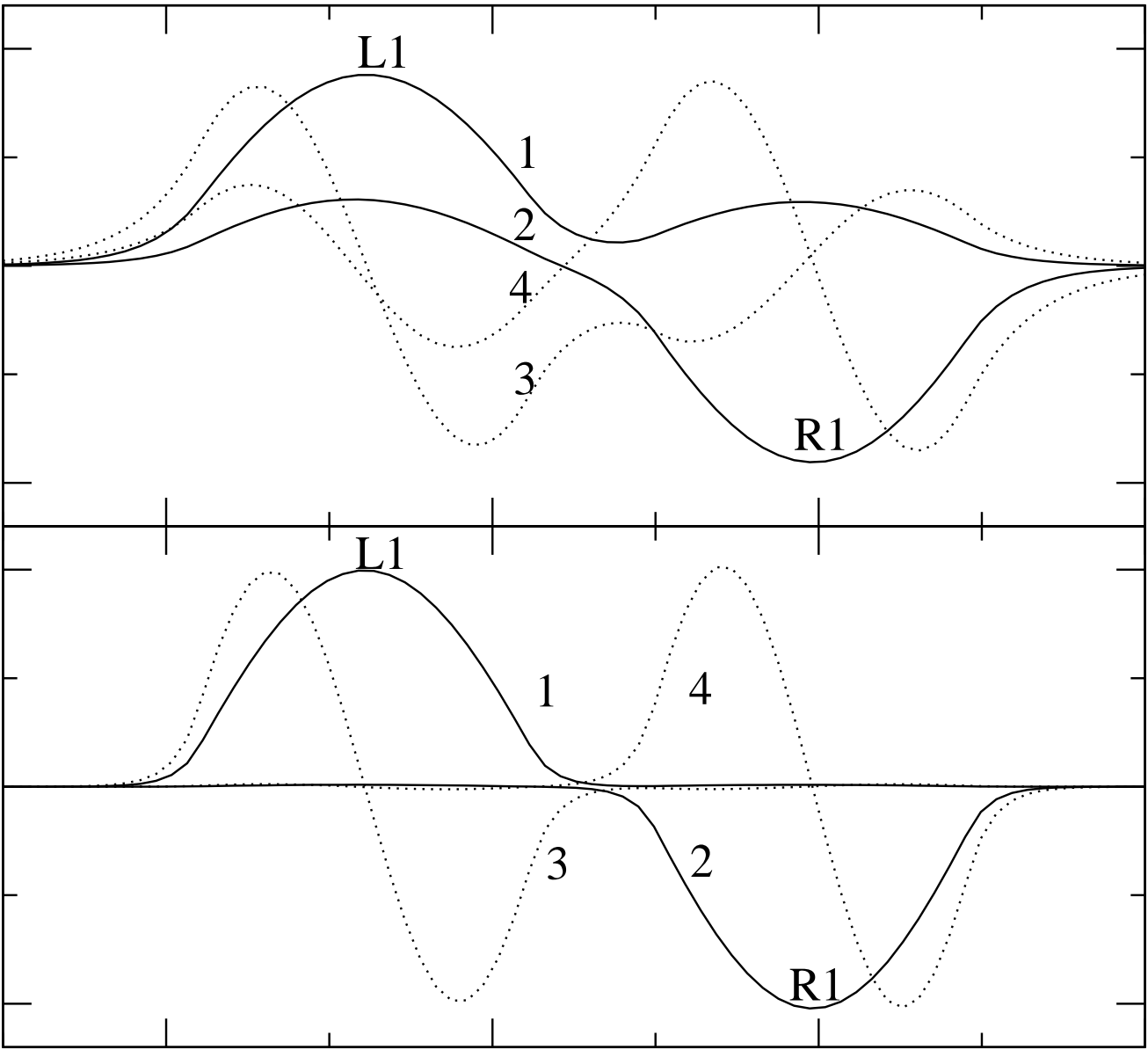
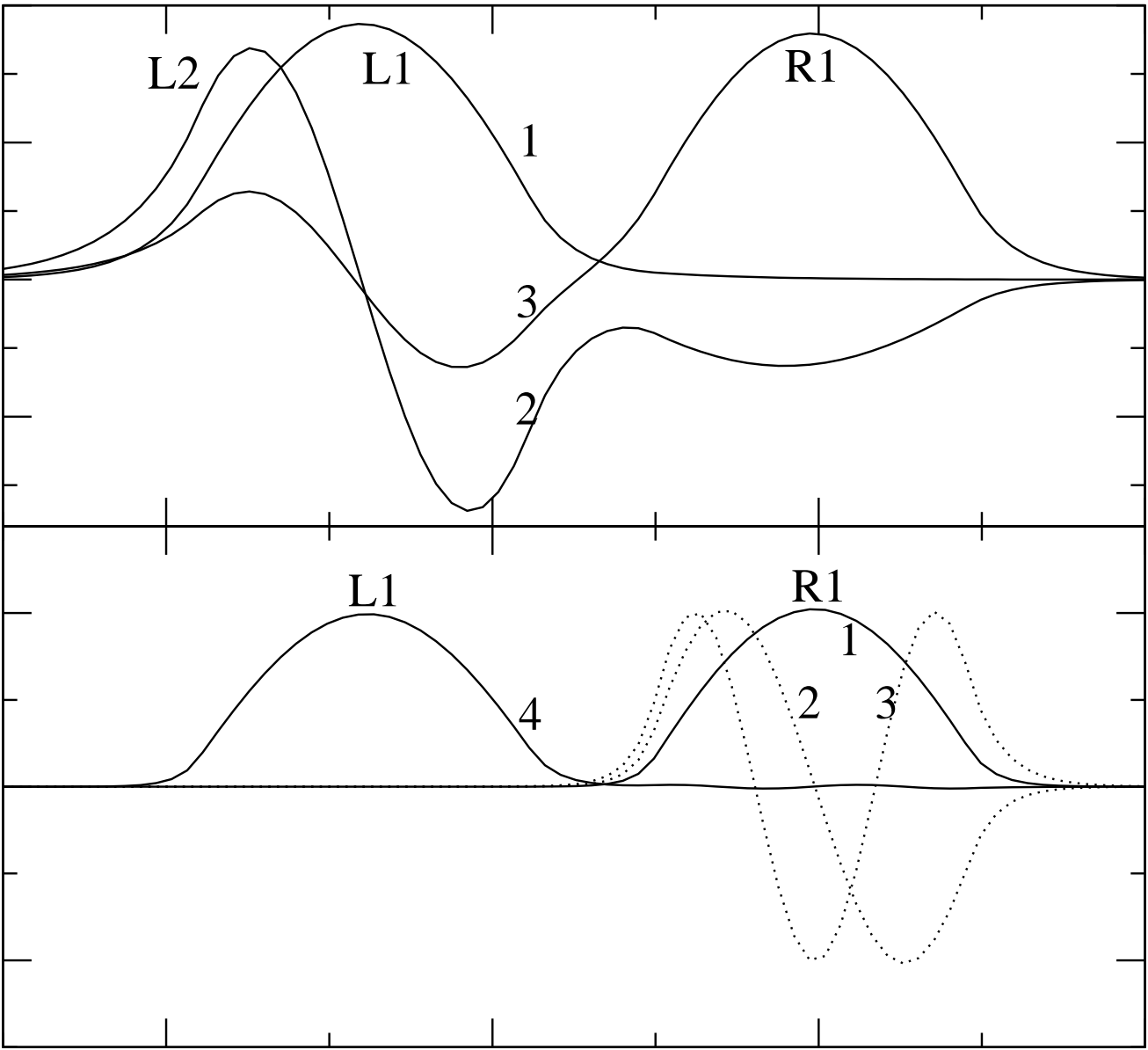
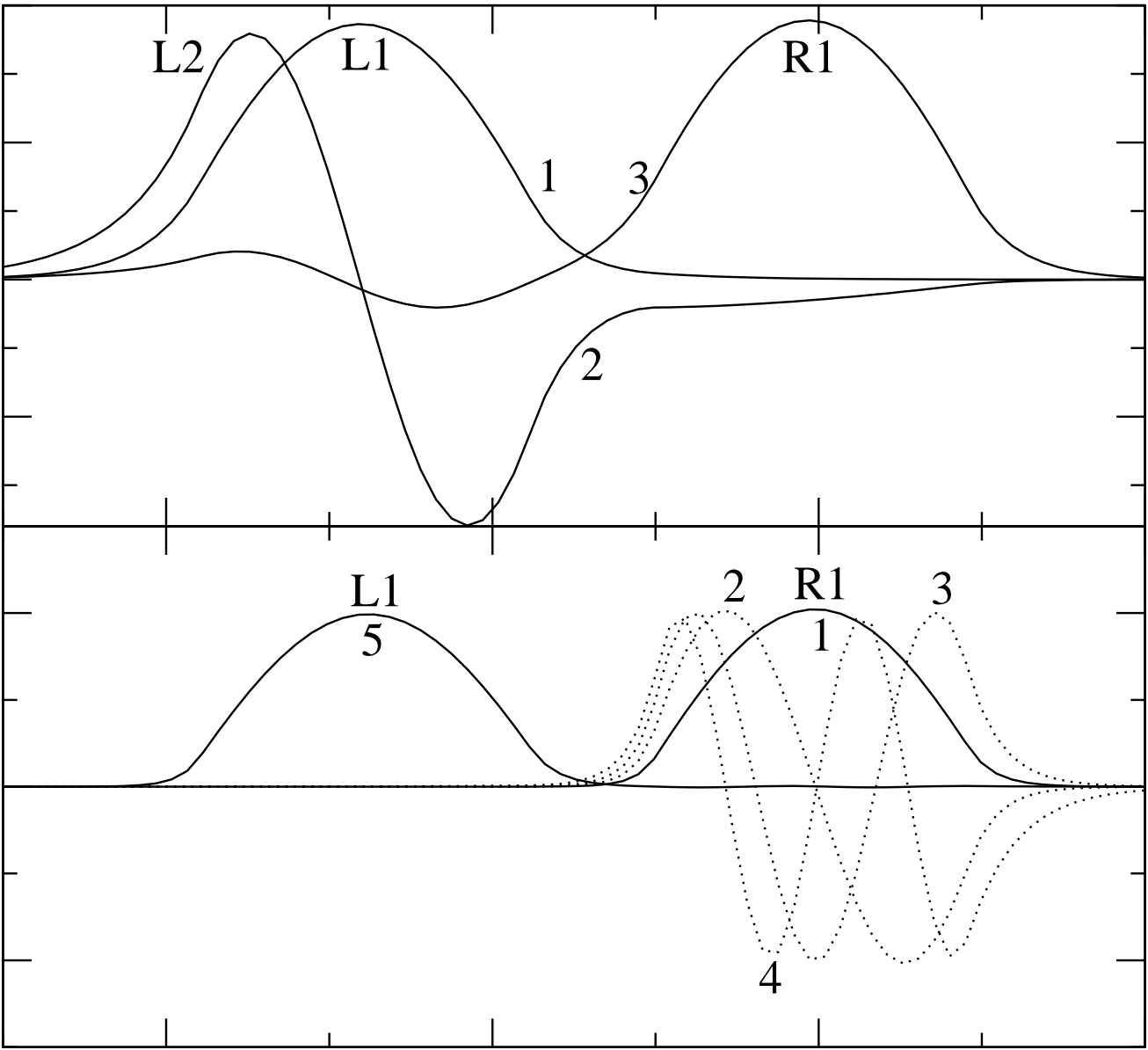
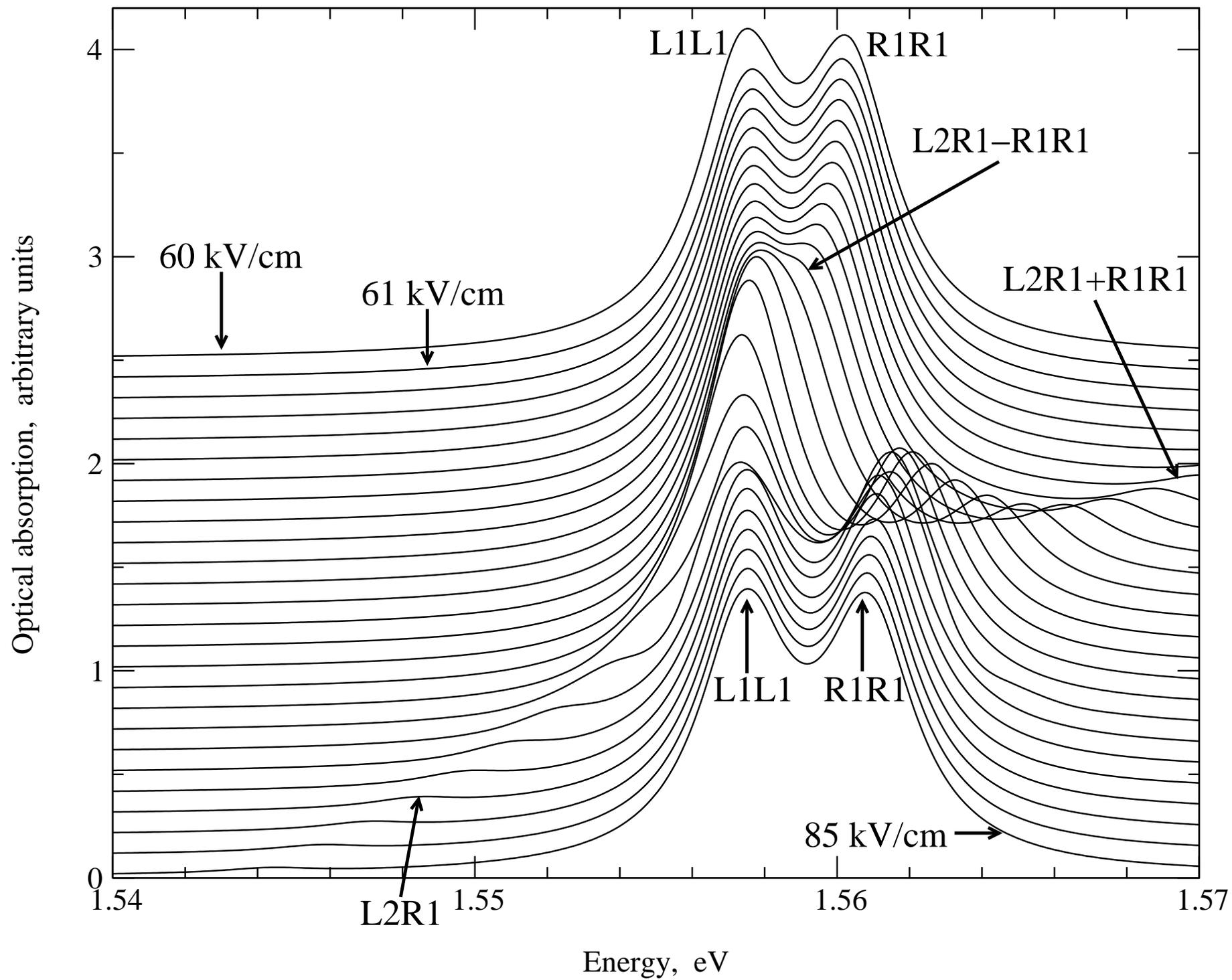
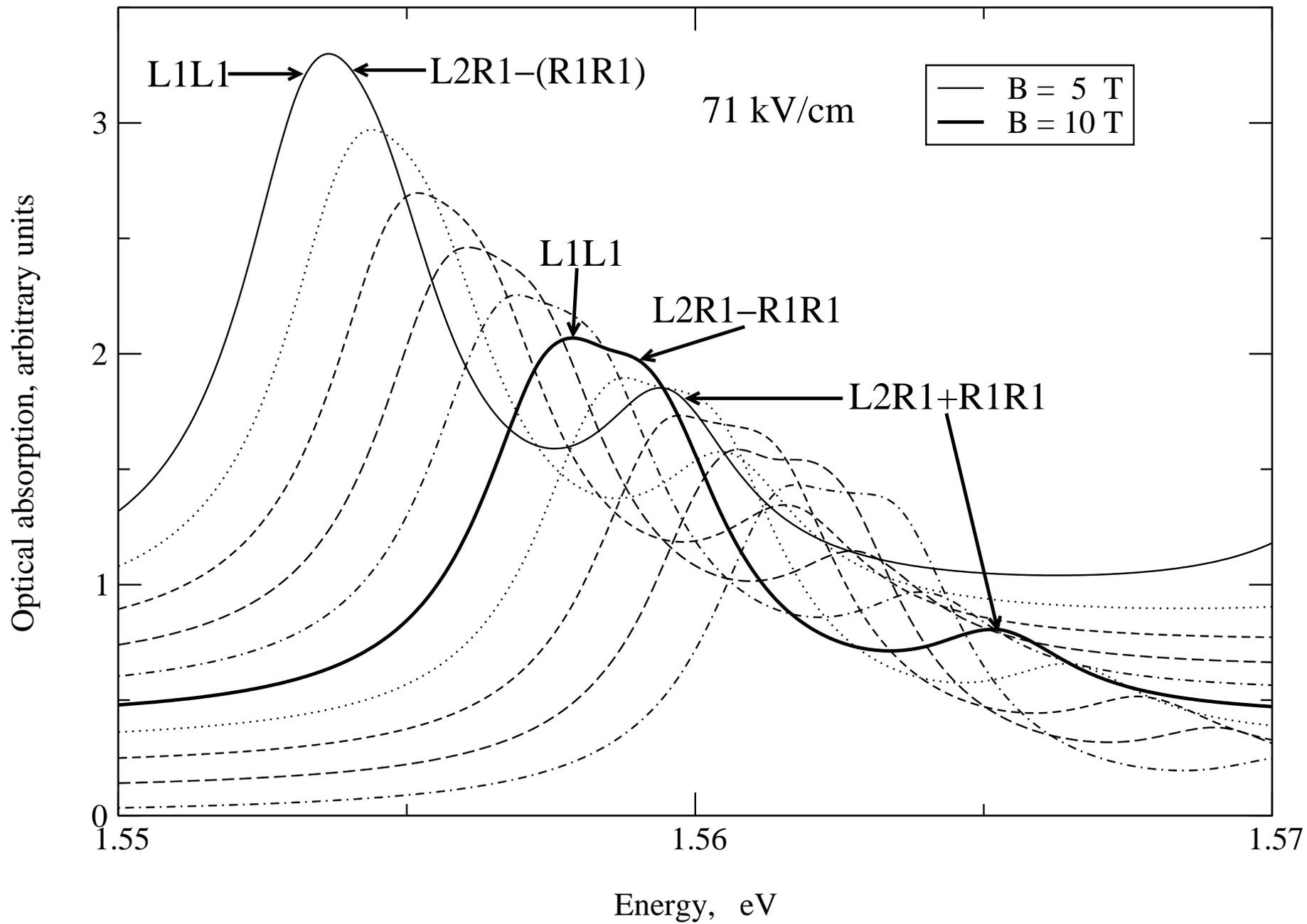
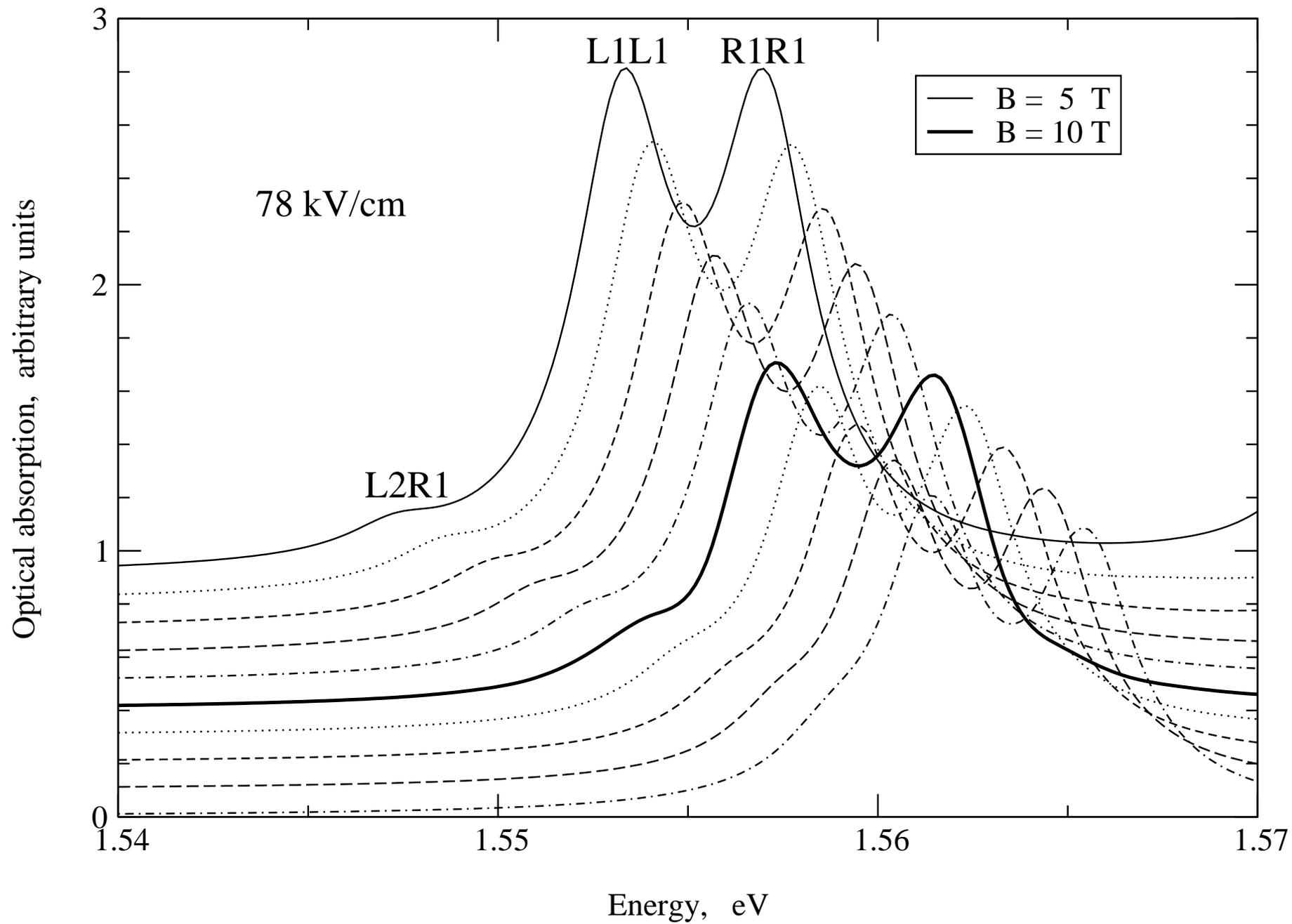
\end{document}